\documentclass[prb,
twocolumn,
showpacs,
preprintnumbers,
amssymb, amsmath, floatfix]{revtex4}
\usepackage{graphicx}

\newcommand{\nn}{\ensuremath{\nonumber}}

\renewcommand{\vec}[1]{\ensuremath{\boldsymbol{\mathrm{#1}}}}

\newcommand{\pd}[2]{\ensuremath{\frac{\partial #1}{\partial #2}}}

\newcommand{\w}{\ensuremath{\omega}}

\begin{document}
\title{Theory of magnetic oscillations in Weyl semimetals}
\author{Phillip E. C. Ashby}
\email{ashbype@mcmaster.ca}
\address{Department of Physics and Astronomy, McMaster University, Hamilton, Ontario, Canada L8S 4M1}

\author{J. P. Carbotte}
\email{carbotte@mcmaster.ca}
\address{Department of Physics and Astronomy, McMaster University, Hamilton, Ontario, Canada L8S 4M1}
\address{The Canadian Institute for Advanced Research, Toronto, Ontario, Canada M5G 1Z8}

\begin{abstract}
Weyl semimetals are a new class of Dirac material that posses bulk energy nodes in three dimensions, in contrast to two dimensional graphene. In this paper, we study a Weyl semimetal subject to an applied magnetic field. We find distinct behavior that can be used to identify materials containing three dimensional dirac Fermions. We derive expressions for the density of states, electronic specific heat, and the magnetization. We focus our attention on the quantum oscillations in the magnetization. We find phase shifts in the quantum oscillations that distinguish the Weyl semimetal from conventional three dimensional Schr\"odinger Fermions, as well as from two dimensional Dirac Fermions. The density of states as a function of energy displays a sawtooth pattern which has its origin in the dispersion of the three dimensional Landau levels. At the same time, the spacing in energy of the sawtooth spike goes like the square root of the applied magnetic field which reflects the Dirac nature of the Fermions. These features are reflected in the specific heat and magnetization. Finally, we apply a simple model for disorder and show that this tends to damp out the magnetic oscillations in the magnetization at small fields.
\end{abstract}

\maketitle
\section{Introduction}

Weyl semimetals are a topological phase of matter and have attracted significant attention as a possible phase in the pyrochlore iridates.\cite{Wan:2011fk,Witczak-Krempa:2012ve} The band structure of a Weyl semimetal consists of pairs of bands crossing at isolated points (Weyl points) in the Brillouin zone (BZ). This is similar to graphene, which has two isolated band touchings in the BZ. Unlike graphene, however, Weyl semimetals are three dimensional. This three dimensionality has important consequences for the stability of these Weyl points. Near the Weyl point the Hamiltonian takes the form 
\begin{align}
\mathcal{H} = \pm v_F \vec{k}\cdot\vec{\sigma},
\end{align}
where $v_F$ is the Fermi velocity, $\vec{\sigma}$ is the vector of Pauli matrices, and $\vec{k}$ is the momentum as measured from the Weyl point. The $\pm$ sign in front characterizes the chirality of the band touching. Since Weyl semimetals are three dimensional, this Hamiltonian uses all three of the Pauli matrices. The Pauli matrices form a basis, and so a perturbation cannot be added to the system that gaps out the Weyl point. The only way these isolated Weyl points can be destroyed is if two Weyl points of opposite chirality meet in the BZ. This stability argument relies on the fact that only two bands touch at each Weyl point. If there is additional degeneracy, then perturbations in this degenerate subspace can be used to gap out the Weyl point. For this reason, the Weyl semimetal phase breaks at least one of time-reversal or inversion symmetry; either of these symmetries would make the band touching at the Weyl point doubly degenerate.

Weyl semimetals are expected to have a variety of novel physical properties. Despite the fact that a Weyl semimetal has a vanishing density of states at $\w=0$ it is expected to have a finite DC conductivity, even at zero temperature.\cite{Hosur:2012fk,Rosenstein:2013fk} In this respect, the Weyl semimetal phase behaves more like a metal than an insulator. For Weyl semimetals with broken time reversal symmetry, there is expected to be an anomalous Hall effect.\cite{Ran2011:fk,Xu:2011kx,Goswami:2012fk,Burkov:2011kx,Zyuzin:2012zr} This anomalous Hall effect is proportional to the separation of Weyl points in momentum space. Weyl semimetals that break inversion symmetry are characterized by a chiral-magnetic effect.\cite{Jian-Hui:2013kx,Chen:2013uq} The chiral-magnetic effect is a dissipationless current that flows parallel to an applied magnetic field and is proportional to the separation of Weyl points in energy. Weyl semimetals also have interesting gapless surface states.\cite{Wan:2011fk,Hosur:2012zr} These states are open segments of Fermi surface called Fermi arcs. The Fermi arc surface states are a direct consequence of the non-trivial topology of the Weyl points. These surface states should be observable by photoemission experiments and would be a clear signature of the Weyl semimetal phase.

Despite the recent interest in Weyl semimetals, one has yet to be unambiguously identified in Nature. However, there are many possible suggestions for where to find the Weyl semimetal phase. The first suggestion was from the family of iridium based pyrochlores.\cite{Wan:2011fk} These pyrochlores have been studied in the context of the metal-insulator transition. Band structure calculations for the pyrochlore Y$_2$Ir$_2$O$_7$ show that it has 24 Weyl points. The pyrochlore iridates, which have inversion symmetry, but break time reversal symmetry, are an exciting candidate for the Weyl semimetal phase. Proposals have been made to engineer a Weyl semimetal using magnetically doped topological insulators,\cite{Liu:2013ly,cho2011:fk} or using topological insulator heterostructures.\cite{Burkov:2011kx,Burkov:2011ys,Zyuzin:2012zr,Halasz:2012ly} First principles calculations also give the possibility of the Weyl semimetal phase in HgCr$_2$Se$_4$\cite{Xu:2011kx} as well as in $\beta$-cristobalite BiO$_2$.\cite{Young:2012vn} Another exciting possibility is that certain quasicrystals may host the Weyl semimetalic state. A recent paper reported that in certain quasicrystaline samples the optical conductivity was linear over a large range of frequencies.\cite{Timusk:2013fk} Combined with the fact that quasicrystals break inversion symmetry, a linear optical conductivity is consistent with that for a Weyl semimetal. Very recently, three dimensional Dirac cones have been observed in Cd$_3$As$_2$\cite{Neupane:fk2013,Borisenko:fk2013} as well as in zinc-blend HgCdTe.\cite{Orlita:2013fk}

Weyl semimetals are predicted to have an interesting response in applied magnetic fields.\cite{vafek:2013fk,Parameswaran:fk2013,Ashby:2013ys} This response is similar to graphene in an applied magnetic field where the Dirac dispersion plays an important role in the physics.\cite{Gusynin:2007vn,Gusynin:2009vn,Pound:2012vn} In this paper we report on magnetic oscillations in a Weyl semimetal. It is important to fully characterize the Weyl semimetal state, so that candidate materials can be properly identified. It is a primary aim of this work to provide predictions for the behavior of various properties of the Weyl semimetal state that can be used by experimentalists to distinguish it from other states of matter. We compute the specific heat and magnetization for a single Weyl point. In a system that has $N_W$ Weyl points, our results should be multiplied by a factor of $N_W$. 

This paper is organized as follows. In section \ref{sec:dos} we calculate the density of states for a Weyl semimetal. We compute the density of states in the finite magnetic field case, and use the Poisson resummation formula to obtain an expression in the small magnetic field limit. The use of the Poisson resummation allows us to extract the form of the magnetic oscillations. In section \ref{sec:comp} we compare our results to that of conventional Schr\"odinger Fermions. We find that the quantum oscillations contain phase information that distinguishes the Weyl semimetal from conventional Schr\"odinger Fermions. In section \ref{sec:spec} we compute the electronic specific heat. The strength of the magnetic field affects the specific heat scaling. Thus, a magnetic field could be used to extract the electronic contribution to the specific heat from the phonon background. We also discuss the cases of finite doping, that is, a finite chemical potential. In section \ref{sec:mag} we compute the quantum oscillations of the magnetization. The vacuum contribution to the magnetization is calculated in appendix \ref{sec:apa}. The oscillations of the magnetization contains phase information similar to that discussed in section \ref{sec:comp}. Finally, we make some concluding remarks in section \ref{sec:conc}. Throughout our paper we work in units where $\hbar = v_F = k_B = 1$.

\section{Density of states}
\label{sec:dos}
We begin by calculating the density of states of a Weyl semimetal in a magnetic field. The density of states plays a crucial role in our calculations of the magnetic oscillations in a Weyl semimetal. The energy levels for an isolated Weyl point in a magnetic field oriented along the $z$ direction take the form
\begin{gather}
E_0 = -k_z,\\
E_{n\lambda} = \lambda\sqrt{2n/l_B^2+k_z^2} = \pm E_n\qquad n\ge1,
\end{gather}
where $l_B^2 = c/eB$ is the magnetic length. The corresponding density of states, $N(\w)$, is given by
\begin{align}
\label{eqn:dos} N(\w) = \frac{1}{8\pi^2l_B^2}\int dk_z \sum_{n=0}^\infty (2n+1)\mathcal{A}(\w,n),
\end{align}
where
\begin{align}
\nn\mathcal{A}(\w,n) &= \delta(\w-E_n)+\delta(\w+E_n)\\
&-\delta(\w-E_{n+1})-\delta(\w+E_{n+1}).
\end{align}
In this form the density of states compactly takes into account that the $n=0$ Landau level has half the degeneracy of all the other levels. Now, we evaluate the integral over $k_z$. There are contributions from
\begin{align}
k_{\pm} = \pm\sqrt{\w^2-\frac{2n}{l_B^2}}.
\end{align}
\begin{figure}
\centering
 \includegraphics[width=\linewidth]{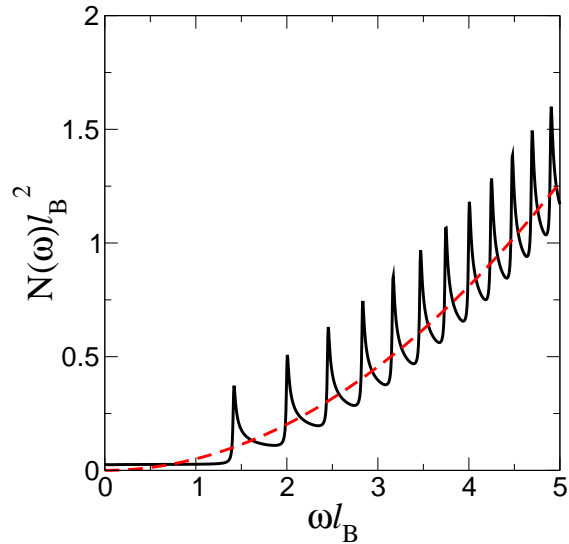}
 \caption{The density of states in magnetic field as a function of frequency. There is a flat piece at low frequency until the magnetic energy is reached. After this there is a set of sharp peaks due to the Landau levels. The 0 field limit is shown in red.}.
 \label{fig:dos1} 
 \end{figure}

\begin{figure}
\centering
 \includegraphics[width=\linewidth]{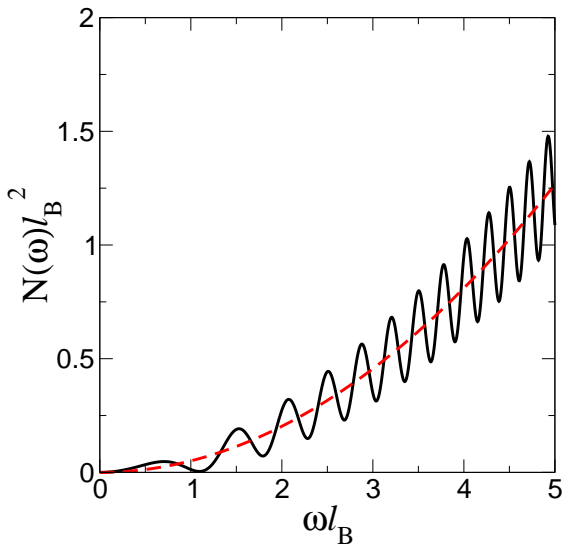}
 \caption{The density of states in the $B\rightarrow 0$ limit as a function of frequency. The dashed red line shows the $B=0$ quadratic density of states, while the solid black line shows the density of states including the first harmonic ($k=1$).}
 \label{fig:dos2} 
 \end{figure}

Performing the integration over $k_z$ in Eq. (\ref{eqn:dos}) gives
\begin{align}
\label{eqn:dos1}N(\w)=\frac{1}{4\pi^2l_B^2}\left\{1+2|\w|\sum_{n=1}^{\left\lfloor \frac{\w^2l_B^2}{2}\right\rfloor}\frac{1}{\sqrt{\w^2-\frac{2n}{l_B^2}}}\right\}.
\end{align}
In the above $\lfloor x\rfloor$ is the integer part of $x$. The density of states, Eq. (\ref{eqn:dos1}) is flat below the magnetic energy scale. Above the magnetic energy there is a series of peaks from the Landau level structure of the Weyl semimetal. These peaks originate from the square root singularity in Eq. (\ref{eqn:dos1}) and physically correspond to the large degeneracy of the Landau levels. 

This is shown in Fig. \ref{fig:dos1} where we plot the universal curve $N(\w)l_B^2$ as a function of $\w l_B$ (Eq. (\ref{eqn:dos1})) and compare it with the $B=0$ density of states which is given by Eq. (\ref{eq:weylback}) and is quadratic in energy. The magnetic oscillations about this background have a sawtooth appearance originating from the $k_z$ dispersion of the Landau levels. If we were dealing with a two dimensional Dirac system (as in graphene) this dispersion would not arise, and the density of states would consist of a series of symmetric peaks. An important feature of the curve in Fig. \ref{fig:dos1} that has its origin in the Dirac nature of the charge carriers involved, is the spacing of the square root singularities goes like the square root of the magnetic field. As we will discuss later, this is distinct from Schr\"odinger fermions for which the spacing of the peaks is linear in the magnetic field. 

In this work we will often employ the Poisson resummation formula. We stress that the Poisson resummation formula is an identity. For a complex function $f(n)$ we use the following form for the resummation formula:

\begin{align}
\nn\frac{1}{2}f(0)+\sum_{n=1}^\infty f(n) =& \int_0^\infty f(x) dx\\
&+ 2 \textrm{Re}\sum_{k=1}^\infty \int_0^\infty f(x) e^{2\pi i k x} dx
\end{align}
We use the Poisson formula on Eq. (\ref{eqn:dos}). The $B$ independent term in the resummation is the only one to survive in the $B\rightarrow 0$ limit and is given by
\begin{align}
\label{eq:weylback} N_{B=0}(\w) = \frac{\omega^2}{2\pi^2}.
\end{align}
As expected, we recover the correct background density of states for a Weyl semimetal. The oscillatory piece of the density of states is given by
\begin{align}
 \label{eq:dosos} N_{\textrm{osc}}(\w) =\frac{\w^2}{2\pi^2}\sum_{k=1}^\infty\int_0^1 \frac{dy}{\sqrt{y}}\cos(\pi k\w^2l_B^2(1-y)).
\end{align}
This integral can be computed by using the definition of the Fresnel sine and cosine integrals:
\begin{align}
\int_0^1\frac{\cos(ay)}{\sqrt{y}}dy = \sqrt{\frac{2\pi}{a}} C\left(\sqrt{\frac{2a}{\pi}}\right)\\
\int_0^1\frac{\sin(ay)}{\sqrt{y}}dy = \sqrt{\frac{2\pi}{a}} S\left(\sqrt{\frac{2a}{\pi}}\right)
\end{align}
Using these we get 
\begin{align}
\nn N_{\textrm{osc}}(\w) = \frac{\w}{\pi^2 l_B}\sum_{k=1}^\infty\frac{1}{\sqrt{2k}}&\left[\cos(\pi k \w^2 l_B^2)C(\sqrt{2k}\w l_B)\right.\\
\label{eq:nosc}&+\left.\sin(\pi k \w^2 l_B^2)S(\sqrt{2k}\w l_B)\right].
\end{align}
The background density of states plus the first harmonic ($k=1$ term) (solid black line) is shown in Figure \ref{fig:dos2} along with the background density of states (dashed red line). Here again, as in Fig. \ref{fig:dos1}, the quantity $N(\w)l_B^2$ plotted as a function of $\w l_B$ is a universal curve, applicable for any value of the magnetic field. On comparing the results in Fig. \ref{fig:dos2} with those in Fig. \ref{fig:dos1} we first note that including only the dominant harmonic in the density of states (i.e. keeping on the $k=1$ term in the sum) produces oscillations that are the same order of magnitude as those found when the complete sum is used. There are, however, important qualitative differences at small values of $\w l_B$. For example, the constant value of $1/4\pi^2$ seen in Fig. \ref{fig:dos1} is not reproduced in detail. In the $\w l_B \rightarrow 0$ limit, the approximate data in Fig. \ref{fig:dos2} goes to zero linearly rather than remaining constant at $1/4\pi^2$. Further, the sharp sawtooth behavior is replaced with peak that are less asymmetric. Of course, if all the terms are retained in the Poisson resummation then the two Figures will agree perfectly.

In this paper we wish to identify features in the properties of a Weyl semimetal under and external magnetic field that can be used to distinguish 3 dimensional Dirac Fermions from the three dimensional Schr\"odinger case. We begin by examining the phase and period of the quantum oscillations defined in Eq. (\ref{eq:nosc}). For our purposes it is sufficient to consider the small $B$ limit. In this limit we can use the asymptotic expansions of $C$ and $S$
\begin{align}
\label{eq:as1}C(u)\sim \frac{1}{2}+\frac{1}{\pi u}\sin(\frac{1}{2}\pi u^2),
\end{align}
and
\begin{align}
\label{eq:as2}S(u)\sim \frac{1}{2}-\frac{1}{\pi u}\cos(\frac{1}{2}\pi u^2),
\end{align}
to obtain the form of the oscillations in the density of states. The oscillatory part is finally given by
\begin{align}
\label{eq:weylosc} N_{\textrm{osc}}(\w) = \frac{\w}{2\pi^2l_B}\sum_{k=1}^{\infty}\frac{1}{k}\sin\left(\pi k \w^2l_B^2+\frac{\pi}{4}\right).
\end{align}

\section{Comparison with Schr\"odinger Fermions}
\label{sec:comp}
It is important to contrast the above results with electrons that obey Schr\"odinger dynamics. Here we compare the differences in the density of states. The discussion here generalizes to the quantities calculated later in the manuscript. For Schr\"odinger electrons the density of states is given by
\begin{align}
N^{\textrm{Sch}}(\w) = \frac{1}{4\pi^2 l_B^2}\int dk_z \sum_{n=0}^{\infty}\delta(\w-\w_n),
\end{align}
with $\w_n = \w_c(n+\frac{1}{2})+\frac{1}{2m} k_z^2$, $\w_c = \frac{1}{ml_B^2}$.

After using the Poisson resummation formula the Schr\"odinger density of states (the equivalent of our Eq. (\ref{eq:weylback}) and Eq. (\ref{eq:weylosc})) becomes $N_{B=0}^{\textrm{Sch}}(\w) = \frac{(2m)^{3/2}\sqrt{\omega}}{4\pi^2}$,

and
\begin{align}
N^{\textrm{Sch}}_{\textrm{osc}}(\w) = \frac{m}{2\pi^2 l_B}\sum_{k=1}^\infty \frac{(-1)^k}{\sqrt{k}}\sin\left(2\pi kl_B^2m\w+\frac{\pi}{4}\right).
\end{align}
At this time we point out some essential differences between the Weyl semimetal and a conventional three-dimensional Schr\"odinger fermions. Besides having very different background densities of states ($\w^2$ vs. $\sqrt{\w}$), there are also differences in the quantum oscillations. Firstly, the Weyl semimetal lacks the phase term $(-1)^k$. This factor introduces a phase shift of $\pi$ between the Weyl and Schr\"odinger cases. This phase shift of $\pi$ is reminiscent of the situation for two-dimensional Fermions. In the two dimensional case the dominant oscillation in the density of states has the form:
\begin{align}
\cos(\pi l_B^2 \w^2)\qquad& \textrm{Dirac}\\
\cos(2\pi l_B^2 m\w-\pi)\qquad& \textrm{Schr\"odinger}
\end{align}
The important difference between these two-dimensional results is again a phase shift of $\pi$. In two dimensions the origin of this phase shift can be traced to a topological Berry phase. Viewed from a semiclassical perspective, the area of allowed orbits, $S(\w)$, is quantized as $S(\w)l_B^2 = 2\pi[n+\gamma]$.

The phase mismatch, $\gamma$, was shown in a recent paper\cite{Fuchs:2010dq,Wright:2013bh} to have two distinct contributions,
\begin{align}
\gamma = \gamma_M -\frac{\gamma_B}{2\pi}.
\end{align}
The first term, $\gamma_M$ is known as the Maslov index and is equal to $1/2$. The second term is Berry's phase, and for two-dimensional Dirac fermions is equal to $\pi$. This exact cancellation is made manifest in physical observables as can be seen from the above two forms of the density of states. We can understand the phase shift of $\pi$ in the three-dimensional case in a similar way. We can think of a Weyl semimetal as a stack of gapped Dirac fermions in $k_z$ space with gap $\Delta = k_z$. The resulting physical quantities are averaged over $k_z$. The Berry phase for gapped Dirac Fermions is given by
\begin{align}
\gamma_B = \textrm{sgn}(\Delta)\pi\left(1-\frac{|\Delta|}{\mu}\right).
\end{align}
The average over $k_z$ cancels pairwise for the layers with a finite gap, since there are contributions from both positive and negative $k_z$. This leaves a remaining shift of $\pi$ from the $k_z=0$ layer that is responsible for the difference between the Weyl and Schr\"odinger cases.
As a last remark we notice that the energy dependence of the two cases is rather different. The Dirac energies enter into the oscillations as $\w^2$, while the Schr\"odinger energy appears linearly. This ensures that each term becomes the area of a closed orbit, $S(\w)$, when viewed in the semiclassical quantization picture.

\section{Specific Heat}
\label{sec:spec}
To calculate the specific heat we start from the formula for the internal energy 
\begin{align}
U = \int_{-\infty}^\infty d\w N(\w) \w f(\w),
\end{align}
where $f(\w) = 1/(\exp(\beta(\w-\mu)+1)$ is the Fermi function, $\mu$ is the chemical potential and $\beta = 1/T$ is the inverse temperature. The specific heat is given by $c =\frac{dU}{dT}$.
The derivative with respect to temperature is easily computed using
\begin{align}
\frac{df}{dT} = \frac{\w-\mu}{4T^2}\textrm{sech}^2\left(\frac{\w-\mu}{2T}\right).
\end{align}
This is valid only when $\mu=0$ (no doping) or when $\mu \gg T$ so that any temperature dependence of the chemical potential can be neglected as a first approximation.

\begin{figure}
\centering
 \includegraphics[width=\linewidth,angle=-90]{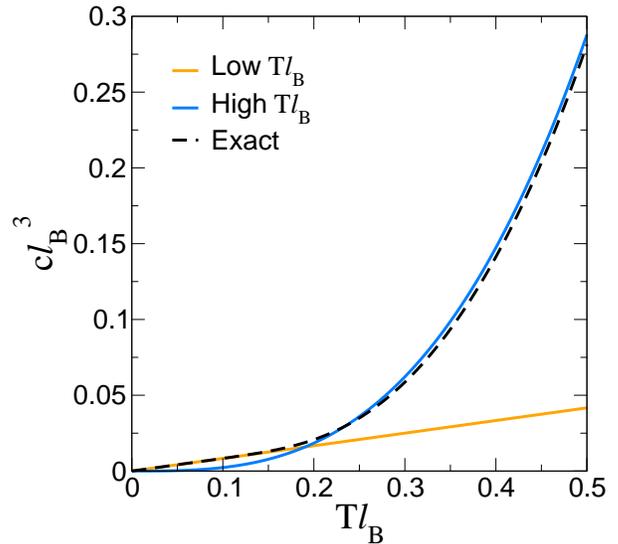}
 \caption{The specific heat as a function of $Tl_B$. Our low $Tl_B$ limit is shown in orange, and the high $Tl_B$ limit is shown in blue. The exact curve (from a direct evaluation of Eq. (\ref{eq:spec}) is shown by the black dashed line. The asymptotic limits agree extremely well with the exact curve.}.
 \label{fig:integral} 
 \end{figure}

We find that the specific heat is given by
\begin{widetext}
\begin{align}
\label{eq:spec}c = \frac{1}{16\pi^2 l_B^2 T^2}\int_{-\infty}^\infty d\w \left[\w+2\sum_{n=1}^\infty \textrm{Re}\left(\frac{\w}{\sqrt{1-\frac{2n}{\w^2l_B^2}}}\right)\right](\w-\mu)\textrm{sech}^2\left(\frac{\w-\mu}{2T}\right)
\end{align}
\end{widetext}
In Figure \ref{fig:integral} we plot the universal function $c l_B^3$ as a function of $T l_B$ for $\mu = 0$. The black dashed line in Figure \ref{fig:integral} is a numerical evaluation of Eq. \ref{eq:spec} with $\mu = 0$. We will compare this curve to the limiting behavious in the two asymptotic regimes ($Tl_B\ll1$ and $Tl_B\gg1$). We first consider the limit where $Tl_B \ll 1$. For $\mu=0$ and $Tl_B \ll 1$ the function $\textrm{sech}^2\left(\frac{\w}{2T}\right)$ is highly peaked around $\omega l_B =0$. In this limit no terms in the sum will contribute. This gives the specific heat
\begin{align}
\label{eqn:specheat} c\sim\frac{1}{12l_B^2} T \qquad Tl_B\ll 1,
\end{align}
i.e. the magnetic scale is much larger than the temperature. This limit is shown as the orange line in Fig. \ref{fig:integral} and fits the exact results perfectly for $Tl_B\lesssim 0.17$. If the chemical potential is finite but still much smaller than the magnetic energy scale ($\mu l_B \ll 1$) then we can still keep just the first term for the specific heat. In this case the low temperature specific heat is given by
\begin{align}
c = \frac{1}{4\pi^2l_B^2}\int_{-\infty}^{\infty}dy (2Ty +\mu)y\textrm{sech}^2(y).
\end{align}
The term proportional to $\mu$ does not contribute to the integral as the integrand is odd. Thus even at finite $\mu$ the specific heat remains equal to the $\mu=0$ case, Eq. (\ref{eqn:specheat}). This is physically expected since the density of states remains constant for the entire range of importance to the specific heat (Figure \ref{fig:dos1}). It is important to remember that this result is only true for both $Tl_B\ll1$ and $\mu l_B \ll 1$. This is the high field limit. This observation should be useful in attempts to identify the Weyl semimetal phase. When $B=0$ the density of states goes like the square of the energy (see Eq. \ref{eq:weylback}) and consequently the specific heat goes like $T^3$. Because the phonon contribution at low temperature also goes like $T^3$ the electronic part cannot be separated out from the phonon background. Applying a large magnetic field changes the electronic heat to a linear in $T$ law that can easily be distinguished from the phonon contribution. However, this does require that the magnetic energy is much larger than the temperature. This is easily accomplished in Dirac materials. Restoring the factor of the Fermi velocity and taking a representative value of $v_F \sim 10^6 m/s$ we get a magnetic energy on the order of $400 K \sqrt{B(T)}$, where $B$ is measured in Tesla. This should be compared to the cyclotron resonance $\w_c \approx 1.4 K B(T) \frac{m_e}{m}$, where $m_e$ is the electron mass. It is clear that the magnetic energy scale is much larger for Dirac Fermions than for the Schr\"odinger case. At high field the low temperature specific heat will remain linear in $T$ and will show no oscillations in a realistic temperature range. This is one of our important results and should help in the experimental identification of a Weyl semimetal.

We now turn to the $Tl_B \gg 1$ limit. In this case we instead use the Poisson resummed version of the density of states. The specific heat is then a sum of two terms, one contribution from the background density of states, and the other from the oscillatory part of the density of states. Keeping the non-oscillatory part of the density of states gives

\begin{align}
c = \begin{cases}
\frac{7}{30}\pi^2 T^3 & \textrm{for } \mu=0, \\
\label{eq:spechigh}\frac{1}{2}\mu^2 T & \textrm{for } \mu \gg T.\end{cases}
\end{align}
The last result is expected for an ordinary electron gas at low temperature. The specific heat is proportional to the electronic density of states at finite chemical potential ($\mu^2$ here). The $(Tl_B)^3$ law of Eq. \ref{eq:spechigh} is shown as the blue line in Fig. \ref{fig:integral}, and matches the numerical results well as long as $Tl_B \gtrsim 0.2$.

Now we examine the oscillatory part
\begin{align}
\nn c_{\textrm{osc}} =& \frac{8T^3}{\pi^2}\sum_{k=1}^\infty\int_{0}^\infty dz z^4\textrm{sech}^2(z)\frac{1}{\sqrt{2k}Tl_B z}\times\\
\nn &\times\left[\cos(4\pi kT^2l_B^2 z^2)C(2\sqrt{2k}Tl_B z)\right.\\
\label{eq:cosc}&\left.+\sin(4\pi kT^2l_B^2 z^2)S(2\sqrt{2k}Tl_B z)\right].
\end{align}
$c_{osc}$ is vanishingly small in the $Tl_B \gg 1$ limit. That is, the only contribution is from the background. This is verified in Fig. \ref{fig:integral} where we compare the exact results to the two asymptotic curves. In the $T l_B \ll 1$ limit, the oscillatory terms add up to give the linear behavior that is correct in this regime (the orange curve in Figure \ref{fig:integral}).

\section{Magnetic oscillations}
\label{sec:mag}

\begin{figure}
\centering
 \includegraphics[width=\linewidth,angle=-90]{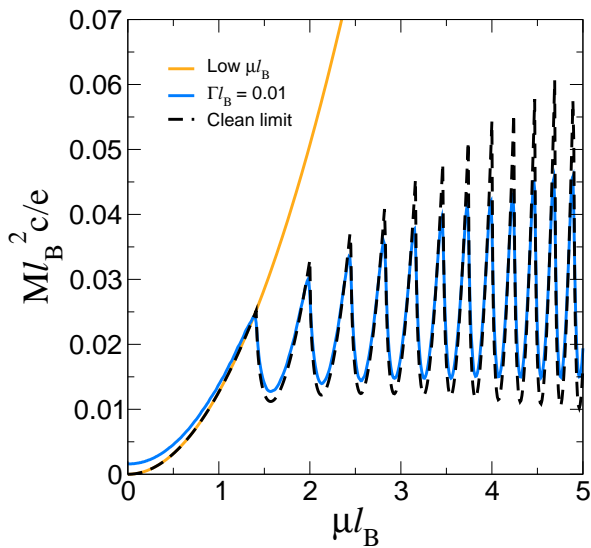}
 \caption{The clean limit of the universal function $\frac{Ml_B^2c}{e}$ as a function of $\mu l_B$ is shown by the black dashed curve. The quantum oscillations of the magnetization onset once $\mu l_B>\sqrt{2}$. The low $\mu l_B$ limit is shown in orange and is quadratic in $\mu l_B$. In blue we show a disorder averaged magnetization with residual scattering rate $\Gamma l_B = 0.01$. The disorder smears the peaks in the quantum oscillations. The disorder also induces additional magnetization in the low $\mu l_B$ region.}
 \label{fig:disorder} 
 \end{figure}

To compute the magnetization for a Weyl semimetal we begin by calculating the thermodynamic potential, $\Omega$. The thermodynamic potential at $T=0$ can be separated into two pieces, a vacuum part, $\Omega_0$, and a part due to quasiparticle excitations, $-W(\mu)$. For more details we refer the reader to section 5 of Sharapov Gusynin and Beck\cite{Sharapov:2004qf}. The zero temperature thermodynamic potential reads
\begin{align}
\Omega(\mu) = \Omega_0 -W(\mu),
\end{align}
where
\begin{align}
W(\mu) = \int_0^\w dx\int_0^x dy N(y).
\end{align}
In Appendix \ref{sec:apa} we show the calculation of the vacuum part. It does not contribute to the magnetic oscillations. We therefore focus on the quasiparticle part, $-W(\mu)$. Also, it is sufficient to compute the thermodynamic potential at zero temperature. To obtain the finite temperature corrections, one can employ the Sommerfeld expansion
\begin{align}
\Omega(\mu, T) = \Omega(\mu)+\frac{\pi^2}{6}T^2\Omega''(\mu)+\frac{7\pi^4}{360}T^4\Omega''''(\mu)+\cdots.
\end{align}
The Sommerfeld expansion can be used as long as the distance between $\mu$ and the nearest Landau level is much greater than the temperature.

The calculation of $W(\mu)$ using Eq. (\ref{eqn:dos1}) is straightforward. We obtain
\begin{align}
\nn W(\mu) = \frac{1}{8\pi^2l_B^2}\left[\mu^2+\frac{2}{l_B^2}\sum_{n=1}^{\left\lfloor \frac{\mu^2l_B^2}{2}\right\rfloor}\left(\mu l_B\sqrt{\mu^2l_B^2-2n}\right.\right.\\
\left.\vphantom{\sum_{n=1}^{\left\lfloor \frac{\mu^2l_B^2}{2}\right\rfloor}}\left.-2n\ln\left(\frac{\mu l_B+\sqrt{\mu^2l_B^2-2n}}{\sqrt{2n}}\right)\right)\right].
\end{align}

The magnetization is then given by $-\pd{\Omega}{B}$. As before, in the $\mu l_B \ll 1$ limit, only the first term in the sum contributes and we have
\begin{align}
M_0(\mu) = \frac{\mu^2e}{8\pi^2c} \qquad \mu l_B \ll 1.
\end{align}
Using the Sommerfeld expansion gives the $T l_B \ll 1$ magnetization $M(\mu = 0, T) = \frac{T^2 e}{24 c}$.
The full magnetization is given by the expression
\begin{align}
\label{eq:mag1}M(\mu) = \textrm{Re}\left(\sum_{n=0}^\infty M_n(\mu)\right)
\end{align}
where $M_n$ is the contribution from the $n$th Landau level:
\begin{align}
\nn M_n(\mu) = \frac{e}{4\pi^2l_B^2c}&\left[\mu l_B\sqrt{\mu^2l_B^2-2n}\right.\\
&\left.-4n\ln\left(\frac{\mu l_B+\sqrt{\mu^2l_B^2-2n}}{\sqrt{2n}}\right)\right].
\end{align}
The quantity $\frac{M l_B^2c}{e}$ is a universal function of $\mu l_B$. A numerical evaluation of Eq. (\ref{eq:mag1}) is shown as the black dashed curve in Figure \ref{fig:disorder}. We see that for high enough fields ($\mu l_B < \sqrt{2}$), there is a complete lack of quantum oscillations. As the field is lowered, the quantum oscillations in the magnetization reveal themselves as a series of peaks. Again, the peak spacing is indicative of the Dirac nature of the bulk electrons. The oscillations occur about a background and the magnitude of the oscillations to the background increases with $\mu l_B$.

To investigate the nature of the oscillations in the large $\mu l_B$ limit we evaluate the thermodynamic potential using the resummed version of the density of states, Eqs.(\ref{eq:weylback})and (\ref{eq:nosc}). Evaluating $W(\mu)$ gives a background term $W_B(\mu) = \frac{\w^4}{24\pi^2}$ and an oscillatory part
\begin{align}
W_{\textrm{osc}}(\mu) = \frac{\mu^4}{12\pi^2l_B^4}\sum_{k=1}^{\infty} \,_2F_3\left(1,1;\frac{5}{4},\frac{7}{4},2;-\frac{\pi^2}{4}k^2\mu^4l_B^4\right),
\end{align}
where $_pF_q(a;b;z)$, the generalized hypergeometric function. From this we obtain the magnetization
\begin{widetext}
\begin{align}
\nn M =& \frac{\mu e}{2\pi^3 l_B c} \sum_{k=1}^{\infty}\frac{1}{(2k)^{3/2}}\left[\cos(\pi k \mu^2l_B^2)S(\sqrt{2k}\mu l_B)-\sin(\pi k \mu^2l_B^2)C(\sqrt{2k}\mu l_B)\right]\\
\label{eq:mag2}&+\frac{\mu^4 l_B^2 e}{6\pi^2c}\sum_{k=1}^{\infty} \,_2F_3\left(1,1;\frac{5}{4},\frac{7}{4},2;-\frac{\pi^2}{4}k^2\mu^4l_B^4\right).
\end{align}
\end{widetext}

We checked that Eq. (\ref{eq:mag2}) is in agreement with Eq. (\ref{eq:mag1}) as long as a sufficient number of terms are kept in the sum over $k$. In the large $\mu l_B$ limit, the first term in Eq. (\ref{eq:mag2}) is the dominant one. This can be seen from the behavior of the ratio of the two terms. In the large $l_B$ limit this ratio behaves like
\begin{align}
\lim_{x\rightarrow\infty} x^3 \,_2 F_3\left(1,1;\frac{5}{4},\frac{7}{4},2;-x^4\right)=0.
\end{align}

\begin{figure}
\centering
 \includegraphics[width=\linewidth,angle=-90]{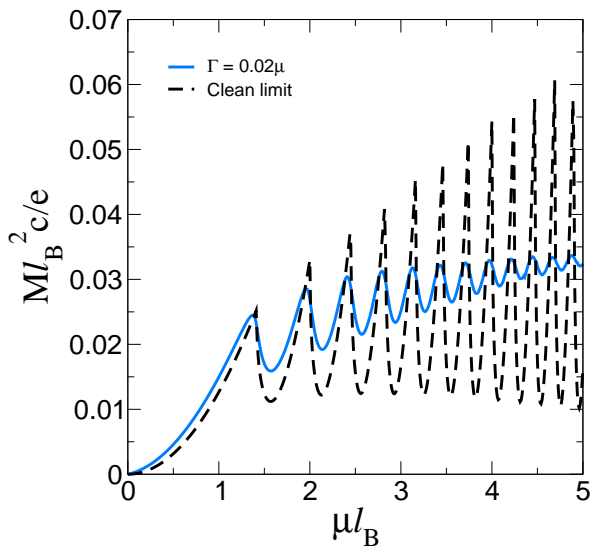}
 \caption{The clean limit of the universal function $\frac{Ml_B^2c}{e}$ as a function of $\mu l_B$ is shown by the black dashed curve. The quantum oscillations of the magnetization onset once $\mu l_B>\sqrt{2}$. In blue we show a disorder averaged magnetization with residual scattering rate $\Gamma = 0.02\mu$. For magnetic fields that are too small (large $l_B$) the oscillations are damped out into the background.}
 \label{fig:scaatter} 
 \end{figure}

We therefore take only the first term, in our discussion of the asymptotic behavior of the magnetization. As we will see, the first term is responsible for the magnetic oscillations.  This limit shows that the amplitude of the oscillations is the dominant contribution in the $\mu l_B \gg 1$ regime. The second term is responsible for the background (of about 0.02 in our Figure \ref{fig:disorder}) about which the oscillations occur. After using the asymptotic expansions for the Fresnel functions we arrive at
\begin{align}
M_{\textrm{osc}} = \frac{\mu e}{8\pi^3l_Bc}\sum_{k=1}^\infty\frac{\cos(k\pi\mu^2l_B^2+\pi/4)}{k^{3/2}},
\end{align}
as the form for the magnetic oscillations. 

We would like to compare this result with that obtained for graphene in the paper by Sharapov, Gusynin and Beck.\cite{Sharapov:2004qf} In their paper they isolated the oscillatory part of the magnetization at large $\mu l_B$. Identifying the gap in graphene as our momentum $k_z$, we can connect our results to theirs by doing an appropriately weighted sum over $k_z$. Averaging their result gives
\begin{align}
\label{eq:shara}\mathcal{M} = -\frac{e}{4\pi^2\mu c}\sum_{k=1}^{\infty}\frac{1}{\pi k}\int_0^\mu dk_z (\mu^2-k_z^2)\sin(\pi k l_B^2(\mu^2-k_z^2)).
\end{align}
After performing the integration, and again using the asymptotic properties of the Fresnel integrals we obtain
\begin{align}
\mathcal{M} = \frac{\mu e}{8\pi^3l_Bc}\sum_{k=1}^{\infty}\frac{\cos(k\pi\mu^2l_B^2+\pi/4)}{k^{3/2}}.
\end{align}
That is, the result is precisely the same as ours. We conclude that the oscillating part of our magnetization is given purely by the first term in Eq. \ref{eq:mag2}. The sum over the hypergeometric functions is responsible for the background that can be seen in Figure \ref{fig:disorder}.  We now comment on the fact that the phase shift of the magnetic oscillations is the same for the magnetization as that discussed in section \ref{sec:comp}.  The identification of this phase shift is a sign of the Dirac nature of the bulk electrons.

Here we have only treated the clean limit. To account for the effect of disorder on the magnetic oscillations, one needs to account for disorder using an appropriate self energy, $\Sigma^{\textrm{imp}}(\w)$. The simplest phenomenological model is to include a constant residual scattering rate, $\Gamma = -\textrm{Im}(\Sigma^{\textrm{imp}}(\w))$. This has the effect of broadening all of the Landau levels by the same amount. In the paper by Sharapov, Gusynin and Beck\cite{Sharapov:2004qf}, they show that this prescription is equivalent to averaging the density of states over a Lorentzian distribution. That is
\begin{align}
N_{\textrm{disorder}}(\w) = \int_{-\infty}^\infty dE \frac{\Gamma}{\pi((E-\w)^2+\Gamma^2)}N_{\textrm{clean}}(\w).
\end{align}
They show that this leads to the appearance of the usual Dingle factors that are responsible for damping out the higher harmonics in the magnetic oscillations. Specifically the Dingle factors appear as $\exp(-2\pi k \mu\Gamma l_B^2)$ in the sum over $k$ in Eq. (\ref{eq:shara}) and thus favor the lowest harmonic ($k=1$ term). The Dingle factors are independent of the graphene gap in their work, and thus will carry over in the same way for a Weyl semimetal. We show a case with residual scattering rate $\Gamma l_B = 0.01$ as the blue curve in Figure \ref{fig:disorder}. The effect of disorder is to broaden the oscillations, and shift up the level at $\mu l_B =0$. Note that scaling $\Gamma$ by the magnetic energy produces a universal function of $\mu l_B$.  A more realistic picture of the effects of scattering is given if we write the scattering rate in terms of the chemical potential. In Figure \ref{fig:scaatter} we show the resulting calculation for a residual scattering rate of $\Gamma = 0.02\mu$ along side the clean limit result.  For quantum oscillations to be visible $\Gamma$ must be smaller than all other energy scales in the problem.  We see that as $\mu l_B$ becomes too large, the disorder averaged magnetic oscillations decrease in amplitude and become hard to resolve from the background.  For the case of a Weyl semimetal where $\mu$ will be a small number, the range of magnetic fields over which one can observe quantum oscillations may be very limited.

In general, disorder leads to more complicated forms of the self energy. The self energy generally has frequency dependence as well as nonzero real and imaginary parts. These details depend on the assumed form of the impurity potential. These complications extend well beyond the scope of the present work.

\section{Conclusion}
\label{sec:conc}
The relativistic Dirac nature of the charge dynamics in Weyl semimetals has important implications for their electronic properties. As an example, the density of states of the charge carriers varies like the square of their energy and is zero at charge neutrality. Applying an external magnetic field, $B$, leads to the formation of Landau levels with spacing proportional to $\sqrt{B}$ rather than the familiar $B$ of Schr\"odinger dynamics. Below the magnetic energy scale the density of states is a finite constant with an inverse square root singular rise at the first Landau level energy. This is followed by a series of successive sawtooth singular rises marking the higher Landau levels. We used the Poisson resummation formula on the density of states and obtained a background density of states that agrees with it's $B=0$ value, on which there are superimposed quantum oscillations. The dominant oscillation varies like $\sin(\pi\w^2l_B^2+\pi/4)$, with a phase shift of $\pi/4$. In this work we related this phase shift to the Berry curvature in a model where a Weyl semimetal is thought of as stacks of gapped graphene layers with gap $\Delta = |k_z|$. It is important to note that this differs from the result for Schr\"odinger fermions by an overall phase shift of $\pi$. This fact could be used to distinguish between Dirac and Schr\"odinger dynamics of the charge carriers. Additionally, the the period could also be used, since the mass $m$ appears in the Schr\"odinger period, but not in the Dirac period.

For temperatures less than the magnetic energy ($Tl_B \ll 1$) the specific heat is linear in $T$ and independent of the doping, provided that the chemical potential remains below the magnetic scale ($\mu l_B \ll 1$ and $T\ll\mu$). In the opposite limit, we take the magnetic scale to be small compared to other scales of interest and found the important contribution is from the background density of states. The background density of states leads to a specific heat that varies like $T^3$ when $\mu = 0$, and has an additive contribution $\propto \mu^2T$ at finite $\mu$. Note that the $T^3$ variation resembles that from the phonon background and cannot be distinguished from it. The electronic contribution can be separated from this background by either doping the system to obtain a finite chemical potential, or by applying a large magnetic field to a $\mu=0$ sample (provided $Tl_B \ll 1$).

We also considered the temperature, magnetic field, and doping dependence of the magnetization, $M$. For $Tl_B \ll1$ we obtain that $M$ has a $T^2$ contribution, as well as a constant piece proportional to the background density of states, $\mu^2$. In the small field limit ($Tl_B \gg 1$) only the oscillatory part of the thermodynamic potential contributes to the magnetization and it varies like $\sqrt{B}\cos(\pi\mu^2l_B^2+\pi/4)$. Again the phase shift of $\pi/4$ is similar to that seen in classical three dimensional systems. This phase shift was compared with the classical systems, and analyzed in terms of the Berry phase.  Lastly, we considered disorder by introducing a residual scattering rate.  We found that at small magnetic fields, the magnetic oscillations become damped out and are are hard to resolve from the background.  This limits the range of magnetic fields over which the quantum oscillations will be observable.

\begin{acknowledgements}
This work was supported by the Natural Sciences and Engineering Research Council of Canada and the Canadian Institute for Advanced Research.
\end{acknowledgements}

\appendix

\section{Calculation of $\Omega_0$}
\label{sec:apa}
\begin{widetext}
Here we compute the contribution from the negative energy states. This contribution is divergent and in this appendix we extract the finite part. To do this we follow the regularization procedure outlined in Bordag {\it et al.}\cite{Bordag:2001fk}. The vacuum contribution to the thermodynamic potential is
\begin{align}
\Omega_0 &= \int_{-\infty}^0 d\w \w N(\w)\\
& = -\frac{1}{8\pi^2 l_B^2}\int_{-\infty}^{\infty}dk_z\left[\sqrt{k_z^2}+2\sum_{n=1}^\infty\sqrt{\frac{2n}{l_B^2}+k_z^2}\right].
\end{align}
For the remainder of the calculation we will include a mass gap, $\Delta$, that we will take to $0$ at the end of the calculation. With the inclusion of the mass gap we have
\begin{align}
\Omega_0 = -\frac{1}{8\pi^2l_B^2}\int_{-\infty}^\infty dk_z\left[\sqrt{k_z^2+\Delta^2}+2\sum_{n=1}^\infty\sqrt{\frac{2n}{l_B^2}+k_z^2+\Delta^2}\right].
\end{align}
Differentiating this gives a divergent magnetization that does not vanish at $B=0$. On physical grounds the magnetization should vanish at $B=0$. We subtract off the $B=0$ divergence to work with the physical magnetization. This leads to the physical contribution to the thermodynamic potential
\begin{align}
\Delta\Omega_0 = -\frac{1}{4\pi^2l_B^2}\int_{-\infty}^{\infty}dk_z\left[\sum_{n=1}^\infty\sqrt{\frac{2n}{l_B^2}+k_z^2+\Delta^2} - \sum_{n=1}^\infty\sqrt{k_z^2+\Delta^2}\right].
\end{align}
To extract the finite contribution we deform the problem by introducing a parameter $s$, that we will take to $0$ at the end. This deformation requires the introduction of a new energy scale, $m$, that serves to give our expression the correct units. We have

\begin{align}
\Delta\Omega_0 = -\frac{m^{2s}}{4\pi^2l_B^2}\int_{-\infty}^{\infty}dk_z\left[\sum_{n=1}^\infty\left(\frac{2n}{l_B^2}+k_z^2+\Delta^2\right)^{\frac{1}{2}-s} - \sum_{n=1}^\infty(k_z^2+\Delta^2)^{\frac{1}{2}-s}\right].
\end{align}
We now make the change of variables $k_z = k_z'\sqrt{\frac{2n}{l_B^2}+\Delta^2}$ (and an analogous change of variables in the second term) so that the integrand factors. We get
\begin{align}
\Delta\Omega_0 = -\frac{m^{2s}}{4\pi^2l_B^2}\int_{-\infty}^{\infty}dk_z'\left[\sum_{n=1}^\infty(1+k_z'^2)^{\frac{1}{2}-s}\left(\frac{2n}{l_B^2}+\Delta^2\right)^{1-s} - \sum_{n=1}^\infty(1+k_z'^2)^{\frac{1}{2}-s}\Delta^{2-2s}\right]
\end{align}
Now the integral over $k_z$ is given by
\begin{align}
\int_{-\infty}^{\infty}(1+k_z'^2)^{\frac{1}{2}-s}dk_z' = \frac{\sqrt{\pi} \Gamma(s-1)}{\Gamma(s-\frac{1}{2})}
\end{align}
The energy now reads
\begin{align}
\Delta\Omega_0 = -\frac{m^{2s}}{4\pi^{\frac{3}{2}}l_B^2}\frac{\Gamma(s-1)}{\Gamma(s-\frac{1}{2})}\left[\sum_{n=1}^\infty\left(\frac{2n}{l_B^2}+\Delta^2\right)^{1-s} - \sum_{n=1}^\infty\Delta^{2-2s}\right].
\end{align}
Now, we recognize the second term as proportional to $\zeta(0) = -1/2$, while for the first we use the representation
\begin{align}
\frac{1}{a^r} = \frac{1}{\Gamma(r)}\int_0^\infty dt t^{r-1}e^{-ta}.
\end{align}
Using this and performing the sum over $n$ leaves
\begin{align}
\Delta\Omega_0 = -\frac{m^{2s}}{4\pi^{\frac{3}{2}}l_B^2}\frac{\Gamma(s-1)}{\Gamma(s-\frac{1}{2})}\left[\frac{1}{2\Gamma(s-1)}\int_0^\infty dt t^{s-2} e^{-t\Delta^2}\left[\coth(teB)-1\right]+ \frac{1}{2}\Delta^{2-2s}\right]
\end{align}
Evaluating the integrals over $t$ leaves us with 
\begin{align}
\Delta\Omega_0 = -\frac{m^{2s}}{4\pi^{\frac{3}{2}}l_B^{4-2s}}2^{1-s}\frac{\Gamma(s-1)}{\Gamma(s-\frac{1}{2})}\left[\zeta\left(s-1,\frac{\Delta ^2l_B^2}{2}\right)+\zeta\left(s-1,1+\frac{\Delta ^2l_B^2}{2}\right)\right].
\end{align}
Now we take $s\rightarrow 0$ and $\Delta\rightarrow 0$. The only ill-behaved part is $\Gamma(s-1)$. As $s\rightarrow 0$
\begin{align}
\Gamma(s-1)\sim -\frac{1}{s}+(\gamma-1)+\mathcal{O}(s)
\end{align}
where $\gamma = 0.577216...$ is Euler's constant. Thus the energy is given by
\begin{align}
\Delta\Omega_0 =-\frac{(eB)^2}{24\pi^2}\left(-\frac{1}{s}+(\gamma-1)+\mathcal{O}(s)\right).
\end{align}
The finite piece is 
\begin{align}
\Delta\Omega_0^{\textrm{finite}} = -\frac{(eB)^2}{24\pi^2}(\gamma-1),
\end{align}
and thus, does not contribute to the magnetic oscillations.
\end{widetext}
\bibliography{biblio}

\end{document}